\documentstyle[12pt]{article}
\begin{document}
\large
\vskip 1cm  

\begin{center}
{\bf      INTERACTION OF RADIATION AND A RELATIVISTIC ELECTRON

      IN MOTIOM IN A CONSTANT MAGNETIC FIELD}
\vskip 0.5cm
           
                  G.M. Filippov
\vskip 0.5cm

I.N. Ul'yanov Chuvash State University, 428015 Cheboksary, Russia

\end{center}
\vskip 1cm

\begin{abstract}
{\bf
This work examines the effect of multiple photon emission on the 
quantum mechanical state of an electron emitting synchrotrotron
radiation and on the intensity of that radiation. Calculations are 
done with a variant of perturbation theory based on the use of extended 
coherent states. A general formula is derived for the number of 
emitted photons, which allows for taking into account their mutual
interaction. A model problem is used to demonstrate the absence 
of the infrared catastrophe in the modified perturbation theory.
Finally, the elecron density matrix is calculated, and the analysis
of this matrix makes it possible to conclude that the degree of
the electron's spatial localization increases with the passage of time
if the electron is being accelerated.}
\end{abstract}

\vskip 2cm

Publised in Zh.\'Eksp.Teor.Fiz.{\bf 113}, 841 (1998) [JETP {\bf 86}, 
459 (1998)].
\newpage

\noindent
{\bf 1. INTRODUCTION}
\par
The effect of radiation on the path of charged particles in a
synchrotron has already been analysed (see, e.g., Ref.1 and the 
literature cited therein). The analysis is based on the classical
Lorenz-Dirac equation or on the solution of a kinetic equation
whose coefficients are the probabilities of quantum transitions
between various stationary states of an electron moving in the 
magnetic field of the synchrotron. Calculations have shown that
in the absence of focusing in the magnetic field there is increase
in the radial fluctuations of the electron path and an increase
in longitudinal fluctuations of the electron's momentum with
the passage of time. In a focusing magnetic field, in the 
initial stages of electron acceleration, the presence of
radiation leads to radiative damping, which damps radial and
vertical oscillations the so-called radiative damping effect).
Lately, research are focused on the analysis of equation of Lorenz-
Dirac type in problem not necessarily related to synchrotron
radiation (see, e.g., Refs. 2---4). New solutions of Lorenz-
Dirac equations have been found for some special cases, and
additional arguments from quantum electrodynamics are invoked
to eliminate nonphysical solutions.

   Despite the indisputable value of the results obtained by
solving Lorenz-Dirac equations, it must be noted that some 
important properties of the states of a particle emitting
radiation, properties that are not directly related to the path
and do not directly influence the above effects --- like an
increase in radial fluctuations of the path --- are excluded
from these results. This is true, in particular, of the evolution
of the particle's wave packet, which affects the radiation and 
hence the radiative friction and the path.

   The present work is an attempt to use a modified perturbation-
theory approach to examine the effect of multiple photon emission
on the evolution of the wave packet of a particle, in particular
when the particle emits synchrotron radiation. Most papers devoted
to the quantum mechanical theory of the synchrotron radiation 
ignore this aspect. It is usually assumed that the particle
emitting radiation has a wave function given by the solution of
the Dirac equation. However, if the emitted radiation is taken 
into account, the particle is only a part of of the quantum
mechanical system and its state cannot be described with the
completeness that is possible in principle in quantum theory.

   A common approach to describing the states of particles that
are members of a large system is to use the concept of the
density matrix.  This work demonstrates that the evolution of
the density matrix suggests that a particle goes with the 
passage of time into states that are more and more localized,
with the particle motion described by the laws of classical
mechanics with ever-increasing accuracy. Thus we have additional
support for the validity of using the Lorenz-Dirac equation
along with a clearer undestanding of the incomleteness of
the physical picture described by this equation.

   The problem of the structure of the wave packets of emitting 
particles is related to the classical model of a distributed
electron studied by Lorenz (see, e.g., Ref. 5). In quantum
electrodynamics this model leads to the well-known problem of
ultraviolet divergence, encountered in the calculations of
the mass, charge, and energy of an elementary particle. 
Furthermore, the renormalization of charge in quantum
electrodynamics reveals the internal inconsistency of the
traditional Feynman formulation of perturbation theory (see,
e.g., Ref. 6). It would be useful to follow the changes in
the difficulties encountered by classical electrodynamics
initiated by changes in the perturbation theory, to establish
which of the above problems is invariant, so to say. It 
might turn out that in the modified theory some of these 
problems can be resolved without resorting to additional 
hypotheses. This might then lead to a new direction in the
development of quantum electrodynamics and the theory of
quantized fields in general. The presen work uses a model to 
show that at least in relation to the infrared catastrophe,
the adopted modification in the theory does not lead to
problems characteristic of the traditional form of the theory.
Calculations are based on a general formula that describes
the mutual interaction of the emitted photons as a 
manifestation of the nonlinearity inherent in quantum 
electrodynamics.

\vskip 0.5cm

\noindent
{\bf 2. EMISSION OF PHOTONS BY A CLASSICAL CHARGED }

\noindent
{\bf PARTICLE}

   We write the Hamiltonian describing the interaction of a
free electromagnetic field and a particle carrying an electric
charge $Z$ (here we use atomic units: $\hbar=1$ and $\vert e\vert=1$):
\begin{equation}\label{1}
\widehat H_{int}=-{1\over c}\int{\bf j}\widehat{\bf A} dV,
\end{equation} 
where the current density vector $\bf j$ is a function of 
coordinates and time. The vector potentual operator is specified
in a three-dimensional transverse gauge,
\begin{equation}\label{2}
\widehat{\bf A}=\sum\limits_{\alpha,\bf q} g_q\lbrace
\widehat f_{\alpha\bf q}{\bf e}_{\alpha\bf q}e^{i{\bf q r}}
+\widehat f^\dagger_{\alpha\bf q}{\bf e}^*_{\alpha\bf q}
e^{-i{\bf q r}}\rbrace
\end{equation}
as a standard linear form in the creation and annihilation
 operators ($\widehat f^\dagger_{\alpha\bf q}$ and $\widehat f
_{\alpha\bf q}$ for photons in states with polarization
$\alpha (\alpha=1,2)$, momentum $\bf q$, and energy $\omega=cq$. 
The polarization vectors ${\bf e}_{\alpha\bf q}$ have unit 
length and are orthogonal to $\bf q$. The coupling constants
$ g_q=(2\pi c^2/\omega\Omega)^{1/2}$ contain the normalization
volume $\Omega$, which does not enter into the final expressions
and thus can be put equal to unity.

   The total Hamiltonian is the sum of the free photon
Hamiltonian
$$
\widehat H_0=\sum\limits_{\alpha\bf q}\omega\hskip 3pt 
\widehat f^\dagger_{\alpha\bf q}\widehat f_{\alpha\bf q}+
const.
$$
and the Hamiltonian (1). We pass to the interaction picture
for field operators, for example,
$${\widehat A}(t)=e^{i{\widehat H}_0 t}{\widehat A}
e^{-i{\widehat H}_0 t}.$$
The equation describing the evolution of the wave vector of
the photon field,  $\vert t)$ in the interaction picture,
\begin{equation}\label{3}
i{d\over dt}\vert t)=\widehat H_{int}(t)\vert t),
\end{equation}
has in the given case an exact solution in the form of the 
direct product of photon coherent states,
\begin{equation}\label{4}
\vert t)=\prod\limits_{\alpha,\bf q} {\rm exp}\lbrack -i\chi_
{\alpha,\bf q}-\widehat f_{\alpha\bf q} Q^*_{\alpha\bf q}+
\widehat f^\dagger_{\alpha\bf q} Q_{\alpha\bf q}\rbrack\vert 
t_0),
\end{equation}
where the initial state vector coinsides, to within the 
arbitrary phase factor, with the vacuum state of the photon 
field:
$$
\vert t_0)=e^{i\phi_0}\vert vac ), \phi_0=const.,
$$
$$ 
Q_{\alpha\bf q}(t)=i{g_q\over c}\int\limits_{t_0}^t dt
^\prime{\bf e}^*_{\alpha\bf q}{\bf j}_{\bf q}(t^\prime)e^
{i\omega t^\prime},
$$
$$
\chi_{\alpha\bf q}(t)=\int\limits_{t_0}^t {\rm Im}[Q_{\alpha \bf q}
(t^\prime)\dot Q^*_{\alpha\bf q}(t^\prime)] dt^\prime,
$$
and $\bf j_{\bf q}(t)$ is the Fourier transform of the current
density.

   Using the exact solution (4), we can calculate all 
quantities of interest. For instance, the mean number of 
photons created by time $t$ is given by
\begin{equation}\label{5}
n^{(0)}_{\alpha\bf q}(t)=\vert Q_{\alpha\bf q}(t)\vert^2.
\end{equation}
Note that this formula yields the mean number of emitted 
photons only as  $t\to\infty$, since creation of a photon 
requires a time interval $c/q$ long, which tends to infinity
as $q\to 0$. In what folows we use the interpretation of
$n_{\alpha\bf q}$. 

   Suppose that the photons are emitted by a point particle
carrying electric charge $Z$ and moving along a path
 ${\bf r}={\bf r}_0(t)$. Then
$$
{\bf j}_{\bf q}(t)=Z{\bf v}_0(t) {\rm exp}[-i{\bf qr}_0(t)],
$$ 
where ${\bf v}_0={\dot{\bf r}}_0(t)$. After summing over
over polarizations we can reduce the time derivative of 
the number of photons (as $t_o\to -\infty$) to the form
\begin{equation}\label{6}
{d\over dt}\sum\limits_{\alpha=1,2} n^{(0)}_{\alpha\bf q}
(t)={Z^2\over c^2}g^2_q\int\limits_{-\infty}^\infty d\tau\lbrack
{\bf v}_0(t-\vert\tau\vert/2+\tau/2){\bf v}_0(t-\vert\tau\vert
/2-\tau/2)-
$$
$$
{1\over q^2}({\bf qv}_0(t-\vert\tau\vert/2+\tau/2))({\bf 
qv}_0(t-\vert\tau\vert/2-\tau/2))\rbrack*
$$ 
$$
\exp \lbrack i\omega\tau-
i{\bf q}({\bf r}_0(t-\vert\tau\vert/2+\tau/2)-{\bf r}_0(t-
\vert\tau\vert/2-\tau/2))\rbrack.
\end{equation}

Applying this equation to the case of synchrotron radiation, we
obtain
\begin{equation}
{d\over dt}\sum\limits_\alpha n_{\alpha\bf q}^{(0)}=
Z^2{v_0^2\over c^2}g^2_q \int\limits_{-\infty}^\infty
d\tau\lbrack \cos\omega_0\tau-{1\over 2}\cos^2\Theta(\cos\omega_0
\tau+\cos((2t-\vert\tau\vert)\omega_0))\rbrack*
$$
$$
\exp \lbrace i\omega\tau-2iqR\, \cos\Theta\,\sin{\omega_0\tau\over 2}\, 
\cos(\omega_0(t-\vert\tau\vert/2))\rbrace,
\end{equation}
where $\omega_0=e{\rm H}_0/\gamma mc, v_0=R\omega_0, {\rm H}_0$ 
is the magnetic field strength, $R$ is the orbit's radius, $m$ is
the particle mass, and $\gamma$ is the Lorenz factor. The angle
$\Theta$ is the inclination of the vector $\bf q$ to the orbital
plane. 

The expression (7) is periodic in time, with period $T_0=2\pi/\omega_0$.
Averaging over one period, we obtain
\begin{equation}\label{8}
{d\over dt}\sum\limits_\alpha{\overline{n^{(0)}_{\alpha\bf q}}}
=Z^2{v_0^2\over c^2} g^2_q\int\limits_{-\infty}^\infty d\tau\,
 e^{i\omega\tau}\Biggl[{1\over 2}\cos^2\Theta\: J_2\left(2qR\, 
\cos\Theta \,\sin{\omega_0\tau\over 2}\right)+
$$
$$
\cos\omega_0\tau \left(1-{1\over 2}\cos^2\Theta\right)\: J_0\left(2qR\, 
\cos\Theta\, \sin{\omega_0\tau\over 2}\right) \Biggr].
\end{equation}

Next, we allow for the fact that for any periodic function
$F(\tau)$,
\begin{equation}\label{9} 
\int\limits_{-\infty}^\infty e^{-i\omega\tau}F(\tau)d\tau=
\sum\limits_{n=-\infty}^\infty e^{in\omega T_0}\int\limits_0^{T_0}
e^{i\omega\tau}F(\tau)d\tau,
\end{equation}
where the sum of exponentials can be transformed into a sum
of delta functions:
$$
\sum\limits_{n=-\infty}^\infty e^{in\omega T_0}={2\pi\over T_0}
\sum\limits_{n^\prime=-\infty}^\infty \delta(\omega-n^\prime
\omega_0).
$$
Combining this with (9), we can transform (8) to the following
form:
\begin{equation}\label{10}
 {d\over dt}\sum\limits_\alpha{\overline{n^{(0)}_{\alpha\bf q}}}
=Z^2{v_0^2\over c^2} g^2_q\sum\limits_{n=0}^\infty{1\over\pi}
\int\limits_0^\pi dx\, e^{2inx}\lbrace \cos2x\, J_0(2qR\, \cos\Theta\, 
\sin x)(1-{1\over 2}\cos^2 \Theta)+
$$ 
$$
{1\over 2} \cos^2\Theta\, J_2(2qR\, \cos\Theta\, \sin x)\rbrace 
2\pi\delta(\omega-n\omega_0).
\end{equation}
Since $\omega=cq>0$, the sum in (10) goes from 0 to $\infty$. 

   Next we have$^7$
$$
\int\limits_0^\pi e^{2i\mu x}\, J_{2\nu}(2a\,\sin x)dx=\pi e^{i\pi\mu} 
J_{\nu-\mu}(a)\, J_{\nu+\mu}(a),
$$
and the recurrence formulas
$$
J_{n+1}(z)+J_{n-1}(z)={2n\over z}J_n(z);
$$
$$
J_{n-1}(z)={n\over z} J_n(z)+J_n^\prime(z);
$$
$$
J_{n+1}(z)={n\over z} J_n(z)-J_n^\prime(z).
$$
As a result, Eq. (10) becomes
\begin{equation}\label{11}
{d\over dt}\sum\limits_\alpha{\overline{n^{(0)}_{\alpha\bf q}}}
=Z^2 g^2_q\sum\limits_{n=0}^\infty 2\pi\delta(\omega-n\omega_0)
\lbrack {\rm tg}^2\Theta\, J_n({nv\over c} \cos\Theta)+
{v^2_0\over c^2}J_n^{\prime^2}({nv\over c} \cos\Theta)\rbrack.
\end{equation}
Equation (11) can be used, in particular, to obtain the well-
known Schott formula. Thus, for the mean intensity of synchrotron
radiation the semiclassical theory yields results that coinside
with classical results. The semiclassical theory provides
additional information (in comparison to that provided by
classical electrodynamics) only in the sence that it makes it 
possible to calculate the fluctiations in the number of the
emitted photons, their mean energy, and total momentum. 
Similar results can be obtained for the case in which a charged 
particle moves along an arbitrary path$^8$. In all 
cases we at classical formulas for the mean intensity of 
radiation emitted by the particle. Moreover, calculations
of the mean electromagnetic field that accompanishes a
charged particle moving {\it in vacuo} lead to well-known
expressionsfor the retarded potentials$^9$. This 
agreement between the semiclassical and classical theories
forms the basis for a more accurate quantum mechanical 
theory of interaction of radiation and an emitting particle.
\vskip 5 pt

\noindent
{\bf 3. QUANTUM MECHANICAL THEORY}

   We consider the interaction of an electron and the 
radiation emitted by that electron. We pass to the furry 
representation and write the wave operator of the electron
in the form of an expansion in the stationary states of 
type (A2) (see the Appendix):
$$
{\widehat \psi}=\sum\limits_{\xi}{\widehat d}_{\xi}\psi_{\xi},
$$
where we have excluded the antiparticle operators, since
allowing for the contribution of particle-antiparticle
intermediate states leads only to small corrections to the 
phenomena considered. The creation and annihilation 
operators, ${\widehat d}^\dagger_\xi$ and ${\widehat d}_\xi$,
must obey the standard Fermi commutation relations. The
current density operator is approximately (without allowing
for electron-positron pair contributions) given by
\begin{equation}\label{12}
{\widehat j}_a(t)=c{\widehat \psi}^\dagger \alpha_a{\widehat\psi}
=c\sum\limits_{\xi \xi^\prime} {\widehat d}^\dagger_\xi {\widehat
d}_{\xi^\prime} \psi^*_\xi({\bf r}) \alpha_a\psi_{\xi^\prime}({\bf r})
e^{i(E_{\xi}-E_{\xi^\prime})t}
\end{equation}
(from now on  $a,b\vert=1,2,3$  label the projections of 
vectors on the Cartesian coordinate axes).

   We construct the operator
\begin{equation}\label {13}
{\widehat{\bf j}}_{\bf q}^{(0)}(t)=Z{\dot{\bf r}}_0({\bf q},t)
e^{-i{\bf qr}_0({\bf q},t)}{\widehat\rho}^{(0)}_{\bf q},
\end{equation}
where ${\bf r}_0({\bf q},t)$  is a vector (which needs to be 
determined) that depends on the momentum transfer  $\bf q$ 
and time $t$, and ${\widehat \rho}_{\bf q}^{(0)}$  is the
"zeroth" density operator at time  $t=0$:
$$
{\widehat\rho}_{\bf q}^{(0)}=\sum\limits_{{\bf k},\sigma}{\widehat 
d}^\dagger_{{\bf k}\sigma}{\widehat d}_{{\bf k+q},\sigma}.
$$
Here  ${\widehat d}^\dagger_{{\bf k}\sigma}$ and ${\widehat d}_
{{\bf k}\sigma}$ are the creation and annihilation operators
for an electron in a state with momentum  $\bf k$  and a 
projection of the electron spin on the $z$ axis that takes 
the values  $\sigma=\pm{1\over 2}$. Note that the operator (13) is 
selected in a form that satisfies the charge conservation law.

   We require that the running mean Fourier transform of
the operator (12) coincide with the expectation value of (13):
\begin{equation}\label {14}
(t\vert{\widehat{\bf j}}_{\bf q}(t)\vert t)={\dot{\bf r}}_0
({\bf q},t) e^{-i{\bf qr}_0 ({\bf q},t)}(t\vert{\widehat\rho}
^{(0)}_{\bf q}\vert t).
\end{equation}
Then ${\bf r}_0({\bf q},t)$  must be approximately equal to
the mean position of the particle at time $t$. We define the
deviation of the current from the "zeroth" value to be
$$
\Delta{\widehat{\bf j}}_{\bf q}(t)={\widehat{\bf j}}_
{\bf q}(t)-{\widehat{\bf j}}^{(0)}_{\bf q}(t).
$$
This deviation will be used to build the interaction
operator in the new representation. The above transformation 
is convenient because the operators (13) commute at different
times:
\begin{equation}\label {15}
\lbrack{\widehat j}^{(0)a}_{\bf q}(t),{\widehat j}^{(0)b}
_{{\bf q}^\prime}(t^\prime)\rbrack =0.
\end{equation}

Using (13), we write the electromagnetic interaction 
operator as a sum of two terms, ${\widehat H}_{int}(t)
={\widehat H}_{int}^{(0)}(t)+{\widehat H}_{int}^{(1)}(t)$, 
where
\begin{equation}\label {16}
{\widehat H}^{(0)}_{int}(t)=-{1\over c}\int{\widehat
{\bf j}}^{(0)}(t){\widehat{\bf A}}(t)dV;
\end{equation}
\begin{equation}\label {17}
{\widehat H}^{(1)}_{int}(t)=-{1\over c}\int\Delta{\widehat
{\bf j}}(t){\widehat{\bf A}}(t)dV.
\end{equation}
Then, by virtue of (15), the equation 
\begin{equation}\label {18}
i{d\over dt}\vert t)={\widehat H}_{int}^{(0)}(t)\vert t)
\end{equation}
has an exact solution in the form of a direct product of
the vectors of extended (or modified, in terminology of
Ref. 10) coherent states,
\begin{equation}\label {19}
\vert t)=\prod\limits_{\alpha,\bf q}\exp \left(-i{\widehat\chi}_
{\alpha\bf q}-{\widehat f}_{\alpha\bf q}{\widehat Q}
^\dagger_{\alpha\bf q}+{\widehat f}^\dagger_{\alpha\bf q}
{\widehat Q}_{\alpha\bf q}\right)\vert 0),
\end{equation}
where (at $t_0=0$)
\begin{equation}\label {20}
{\widehat Q}_{\alpha\bf q}(t)=i{g_q\over c}\int\limits_0
^t dt^{\prime} {\bf e}^*_{\alpha\bf q}{\widehat{\bf j}}^{(0)}_
{\bf q}(t^{\prime})e^{i\omega t^{\prime}},
\end{equation}
\begin{equation}\label {21}
{\widehat\chi}_{\alpha\bf q}(t)=-{i\over 2}\int\limits_
0^t\lbrace{\widehat{\dot Q}}^\dagger_{\alpha\bf q}(t^{\prime})
{\widehat Q}_{\alpha\bf q}(t^{\prime})-{\widehat Q}_{\alpha\bf q}
^\dagger (t^{\prime}){\widehat{\dot Q}}_{\alpha\bf q}(t^{\prime})
\rbrace dt^{\prime}.
\end{equation}
The initial state vector $\vert 0)$ is the direct product of
the vacuum state of the electromagnetic field, $\vert vac)$, 
and the vector of the initial state of the moving particle,
 $\vert\phi)$, described by the wave function $\phi({\bf r}),
i.e., \: \vert 0)=\vert\phi,vac)$. 

   We have chosen $t_0=0$ to be zero rather than $-\infty$
due to the fact that, as further calculation show, the
temporal sequence of changes in the state of a moving particle
that interacts with the field of the radiation it emits is
highly important. In this approach there are sure to be 
problems assosiatad with the inteaction turning on, which 
violates the charge conservation, and with the generation 
of virtual radiation, which is the consequence of such 
violation. To avoid the need to discard fictitious terms,
one can resort to turning the interaction on slowly by
replacing the constant $Z$ with a slowly increasing charge
$Z(1-e^{-\epsilon t})$, where  $\epsilon$ is small. The
charge buildup time  $\tau_{in}=\epsilon^{-1}$ must be long
compared to  $\omega^{-1}$, but short compared to the
observation time (here $t$ must be much longer than $\omega^{-1}$). 
After we establish a method for evaluating the integrals
for some definite value of $q$, we can extend it to any
other value of $q$.

   If we ignore the corrections generated by ${\widehat H}_
{int}^{(1)}$, Eq. (19) fully solves the problem of calculating
the physical quantities of interest. In particular,instead of
(5) we have
$$
n_{\alpha\bf q}(t)=(t\vert {\widehat f}_
{\alpha\bf q}^\dagger (t){\widehat f}_{\alpha\bf q}(t)\vert t),
$$
which at ${\bf r}_0({\bf q},t)={\bf r}_0(t)$ leads to a
result coinciding with (5). Thus, if we ignore 
${\widehat H}_{int}^{(1)}$, the current variant of the quantum 
mechanical theory differs from the semiclassical one in
calculations of the mean number of the emitted photons or the 
energy of these photons only when  ${\bf r}_0({\bf q},t)$ 
differs from ${\bf r}_0 (t)$. 

At the same time, corrections due to ${\widehat H}_{int}^{(1)}$, 
can be obtained for any convenient choise of the vectors
${\bf r}_0({\bf q},t)$. Let us put ${\bf r}_0({\bf q},t)={\bf r}_0(t)$.
Then ${\widehat Q}_{\alpha\bf q}(t)=Q_{\alpha
\bf q}(t){\widehat\rho}_{\bf q}$, where $Q_{\alpha\bf q}(t)$ 
is specified by its semiclassical expression
\begin{equation}\label {22}
 Q_{\alpha\bf q}(t)=i{Z\over c}g_q\int\limits_0
^t dt^{\prime} {\bf e}^*_{\alpha\bf q}{\bf v}_0(t^\prime)
\exp \lbrace i\omega t^{\prime}-i{\bf qr}_0(t^\prime)\rbrace.
\end{equation}
To construct a new "modified" perturbation theory in
${\widehat H}_{int}^{(1)}$, we introduce the zero-order evolution
operator 
$$ 
{\widehat U}_0 (t)=\exp \lbrace\sum\limits_{\alpha,\bf q}
{\widehat Q}_{\alpha\bf q}(t) {\widehat f}_{\alpha\bf q}^\dagger
-{\widehat Q}_{\alpha\bf q}^\dagger (t) {\widehat f}_{\alpha\bf q}
-i {\widehat{\chi}}_{\alpha\bf q}(t)\rbrace.
$$
Then (19) can be written as  $\; \vert t)={\widehat U}_0 
(t)\vert 0)$. We also introduce a new representation of 
operators:
\begin{equation}\label {23}
{\widetilde A}(t)={\widehat U}_0^\dagger (t){\widehat A}
(t) {\widehat U}_0 (t).
\end{equation}
The state vector $\vert t>$ in this representation obeys the
equation
\begin{equation}\label {24}
i {d\over dt}\vert t>={\widetilde H}_{int}^{(1)} (t)\vert t>.
\end{equation}
Allowing for (24), we can reduce the expression for the mean
number of photons to
\begin{equation}\label {25}
n_{\alpha\bf q}(t)=n_{\alpha\bf q}^{(0)}(t)+\sum\limits_
{n=1}^\infty (-i)^n\int\limits_0^t dt_1\int\limits_
0^{t_1} dt_2...\int\limits_0^{t_{n-1}} dt_n*
$$
$$
<0\vert \lbrack...\lbrack\lbrack{\widetilde f}_{\alpha\bf q}
^\dagger(t){\widetilde f}_{\alpha\bf q}(t),{\widetilde H}_{int}
^{(1)}(t_1)\rbrack,{\widetilde H}_{int}^{(1)}(t_2)\rbrack,
...,{\widetilde H}_{int}^{(1)}(t_n)\rbrack\vert 0>.
\end{equation}
\vskip 0.5cm

\noindent
{\bf 4. CALCULATING CORRECTIONS IN THE MODIFIED}
{\bf THEORY}

   Writting the series in Eq. (25) explicitly, we find that the
expansion contains terms proportional to even powers of $Z$.
We collect the leading terms of this type, which contain $Z^2$
as a pre-exponential factor. Such terms exist in $ n^{(0)}_{\alpha\bf 
q}(t)$ and in the first and second of the series (25). Note that
here $Z$ is also contained in the exponents entering into ${\widehat U}_0$
and $ {\widehat U}_0^\dagger$. We calculate the first commutator
in (25) via the following auxiliary formulas:
\begin{equation}\label {26}
{\widetilde f}_{\alpha\bf q}(t)=\left({\widehat f}_
{\alpha\bf q}+{\widehat Q}_{\alpha\bf q}(t)\right)
e^{-i\omega t};\hskip 5pt
{\widetilde f}_{\alpha\bf q}^\dagger (t)=\left({\widehat f}_
{\alpha\bf q}^\dagger+{\widehat Q}_{\alpha\bf q}^\dagger 
(t)\right) e^{i\omega t};
$$
$$ 
\lbrack{\widehat f}_{\alpha\bf q},{\widehat U}_0 
(t)\rbrack={\widehat U}_0(t){\widehat Q}_{\alpha\bf q}
(t);\hskip 5pt 
\lbrack{\widehat f}_{\alpha\bf q}^\dagger,{\widehat
U}_0^\dagger (t)\rbrack=-{\widehat U}_0^\dagger (t){\widehat
Q}_{\alpha\bf q}^\dagger (t);
$$ 
$$ 
\lbrack{\widehat f}_{\alpha\bf q}^\dagger,
{\widehat U}_0 (t)\rbrack={\widehat U}_0(t){\widehat Q}
_{\alpha\bf q}^\dagger (t);\hskip 5pt 
\lbrack{\widehat f}_{\alpha\bf q},{\widehat
U}_0^\dagger (t)\rbrack=-{\widehat U}_0^\dagger (t){\widehat
Q}_{\alpha\bf q} (t).
\end{equation}
We put
$$
{\widehat B}_{\alpha\bf q}(t)={\bf e}_{\alpha\bf q}
\Delta{\widehat{\bf j}}_{-\bf q}(t),
$$
so that
\begin{equation}\label {27}
{\widehat H}_{int}^{(1)}(t)=-{Z\over c}\sum\limits_{\alpha,
{\bf q}} g_q\left({\widehat f}_{\alpha\bf q}{\widehat B}_
{\alpha\bf q}(t) e^{-i\omega t}+
{\widehat f}_{\alpha\bf q}^\dagger {\widehat B}_
{\alpha\bf q}^\dagger(t) e^{i\omega t}\right).
\end{equation}
Using (26), we can perform th following transformation:
\begin{equation}\label {28}
\lbrack{\widetilde f}^\dagger_{\alpha\bf q}(t){\widetilde
f}_{\alpha\bf q}(t),{\widetilde H}_{int}^{(1)}(t)\rbrack=
$$
$$ 
{\widehat U}_0^\dagger (t_1)\lbrack \left({\widehat f}
^\dagger_{\alpha\bf q}+{\widehat Q}^\dagger_{\alpha\bf q}(t,t_1)
\right)\left({\widehat f}_{\alpha\bf q}+{\widehat Q}_{\alpha\bf q}
(t,t_1)\right),{\widehat H}_{int}^{(1)}(t_1)\rbrack {\widehat U}_
0(t_1),
\end{equation}
where ${\widehat Q}_{\alpha\bf q}(t,t_1)={\widehat Q}_{\alpha
\bf q}(t)-{\widehat Q}_{\alpha\bf q}(t_1)$. Since the operators
${\widehat Q}_{\alpha\bf q}(t)$ already contain $Z$ as a factor,
the leading terms emerge as a result of the commutation of the
photon operators and ${\widehat H}_{int}^{(1)}$:
\begin{equation}\label {29}
\lbrack{\widetilde f}^\dagger_{\alpha\bf q}(t){\widetilde f}
_{\alpha\bf q}(t),{\widetilde H}^{(1)}_{int}(t_1)\rbrack \cong
{Z\over c}g_q{\widehat U}_0^\dagger(t_1)\biggl( e^{-i\omega t_1}
{\widehat B}_{\alpha\bf q}(t_1)({\widehat f}_{\alpha\bf q}+
{\widehat Q}_{\alpha\bf q}(t,t_1))-
$$
$$
e^{i\omega t_1}({\widehat f}^\dagger_{\alpha\bf q}+
{\widehat Q}^\dagger_{\alpha\bf q}(t,t_1))
{\widehat B}^\dagger_{\alpha\bf q}(t_1)
\biggr) {\widehat U}_0(t_1).
\end{equation}
If we average (29) over the initial state of the system by employing
the equalities
$$ 
{\widehat f}_{\alpha\bf q}{\widehat U}_0(t_1)\vert 0)=
{\widehat Q}_{\alpha\bf q}(t_1){\widehat U}_0(t_1)\vert 0);
$$
$$ 
(0\vert {\widehat U}_0^\dagger(t_1){\widehat f}^\dagger
_{\alpha\bf q}=(0\vert {\widehat U}_0^\dagger (t_1){\widehat
Q}^\dagger_{\alpha\bf q}(t_1),
$$
we get
\begin{equation}\label {30}
(0\vert\lbrack{\widetilde f}^\dagger_{\alpha\bf q}(t){
\widetilde f}_{\alpha\bf q}(t),{\widetilde H}^{(1)}_{int}(t_1)
\rbrack\vert 0)={Z\over c}g_q (0\vert {\widehat U}_0^
\dagger (t_1)\lbrace {\widehat B}_
{\alpha\bf q}(t_1){\widehat Q}_{\alpha\bf q}(t)e^{-i\omega t_1}-
$$
$$
e^{i\omega t_1}{\widehat Q}^\dagger_{\alpha\bf q}(t)
{\widehat B}^\dagger_{\alpha\bf q}(t_1)\rbrace {\widehat U}_0(t_1)
\vert 0).
\end{equation}

   In calculating the next corrections in (25) we emmediately
discard terms that contain pre-exponential factors with $Z$ raized
to a power greater than two. This means that when we plug such 
terms into the second and subsequent terms of the sum in (25)
into the expression for the first-order commutator, we can 
immediately discard terms containing the operators  ${\widehat Q}_
{\alpha\bf q}$ and ${\widehat Q}_{\alpha\bf q}^\dagger$. In the
resulting expressions, the operators  ${\widehat f}_{\alpha\bf q}$
and ${\widehat f}_{\alpha\bf q}^\dagger$ can be freely interchanged
with the operators  ${\widehat U}_0$ and ${\widehat U}_0^\dagger$, 
since their commutators contain heigher-order corrections in $Z$,
which we have just discarded. 

   In view of this, all terms in which the annihilation operators
${\widehat f}_{\alpha\bf q}$ are to the right of other ${\widehat f}$-
operators, or in which the creation operators ${\widehat f}^\dagger_
{\alpha\bf q}$ are to the left of other ${\widehat f}$-operators, must be 
dropped. In the remaining terms the operator products ${\widehat f}_
{\alpha\bf q} {\widehat f}^\dagger_{\alpha^\prime {\bf q}^\prime}$ must 
be replaced by the commutators  $\delta_{\alpha\alpha^\prime} \Delta({\bf q}-
{\bf q}^\prime)$. By performing these transformations we reduce the leading 
terms that appear when we write the double commutator on the right hand side 
of Eq.(25) explicitly to the form
\begin{equation}\label{31}
 -{Z^2\over c^2} g_q^2\biggl({\widehat U}_0^\dagger 
(t_1) e^{-i\omega (t_1-t_2)} {\widehat B}_{\alpha\bf q}
(t_1){\widehat U}_0 (t_1){\widehat U}_0^\dagger (t_2)
{\widehat B}^\dagger_{\alpha\bf q}(t_2)
{\widehat U}_0(t_2)+
$$ 
$$
{\widehat U}_0^\dagger (t_2)
{\widehat B}_{\alpha\bf q}(t_2){\widehat U}_0(t_2)
{\widehat U}_0^\dagger (t_1){\widehat B}^\dagger_{\alpha\bf q}(t_1)
{\widehat U}_0(t_1) e^{i\omega(t_1-t_2)}\biggr).
\end{equation}
Collecting all terms of the specified order,we get
\begin{equation}\label{32}
n_{\alpha\bf q}(t)=\vert Q_{\alpha\bf q}(t)\vert^2-
i{Z\over c}g_q\int\limits_0^t dt_1\biggl( 0\vert{\widehat U}
_0^\dagger(t_1) \lbrace e^{-i\omega t_1}{\widehat B}_{\alpha\bf q}(t_1)
{\widehat Q}_{\alpha\bf q}(t)-
$$ 
$$ 
e^{i\omega t_1}{\widehat Q}^\dagger_{\alpha\bf q}(t) 
{\widehat B}^\dagger_{\alpha\bf q}(t_1)\rbrace {\widehat U}_
0(t_1)\vert 0\biggr)+ {Z^2\over c^2}g_q^2\int\limits_0^t dt_1\int\limits
_0^{t_1} dt_2\biggl( 0\vert{\widehat U}_0^\dagger
(t_1)e^{-i\omega(t_1-t_2)}*
$$
$$
{\widehat B}_{\alpha\bf q}(t_1){\widehat U}_0 (t_1)
 {\widehat U}_0^\dagger(t_2)B^\dagger_{\alpha\bf q}(t_2)
{\widehat U}_0(t_2)+
$$
$$
{\widehat U}_0^\dagger(t_2) {\widehat B}^\dagger_
{\alpha\bf q}(t_2){\widehat U}_0(t_2) {\widehat U}_0^\dagger(t_1)
e^{i\omega(t_1-t_2)}{\widehat B}_{\alpha\bf q}(t_1)
{\widehat U}_0 (t_1)\vert 0\biggr).
\end{equation}
If we now write all terms in (32) that appear because of
plugging the explicit expressions for ${\widehat B}_{\alpha\bf q}(t)$
into (32), collect like terms, and do the necessary canceling,
we arrive at the final result:
\begin{equation}\label{33}
 n_{\alpha\bf q}(t)=
{Z^2\over c^2} g_q^2\int\limits_0^t dt_1\int\limits_0^t dt_2 
\Bigl(0\vert {\widehat U}_0^\dagger (t_1)*
$$
$$
{\bf e}_{\alpha\bf q}{\widehat{\bf j}}^\dagger_{\bf q}(t_1)
{\widehat U}_0(t_1){\widehat U}_0^\dagger(t_2)
{\bf e}^*_{\alpha\bf q}{\widehat{\bf j}}_{\bf q}(t_2)
{\widehat U}_0(t_2)\vert 0\Bigr)e^{-i\omega(t_1-t_2)}.
\end{equation}
Note that at deriving (33), we did not take advantage of the
fact that ${\bf r}_0 ({\bf q},t)$ is independent of $\bf\bf q$,
with the result that the formula still holds in the general
case, in which  ${\bf r}_0(t)$ is replaced by ${\bf r}_0({\bf q},t)$
in(22). 

\vskip 0.5cm

\noindent
{\bf 5. NUMBER OF PHOTONS}

   We assume that in the expansion of the initial state vector
of the particle, $\vert 0)$, the expancion coefficients
$ C_{{\bf k}_i}$ in states with definite momentum  $\vert{\bf k}_i)$ 
have a peak at ${\bf k}_0$, and decrease as ${\bf k}_i$ ®â ${\bf k}_0$ 
devates from ${\bf k}_0$, by the Gauss law
$$
C_{{\bf k}_i}=(2\pi\delta_\bot^2)^{1/2}(2\pi\delta_l^2)^{1/4}
{\rm exp}[-p_i^2\delta_l^2/4-({\bf k}_{i\bot}-{\bf k}_{0\bot})^2
\delta_\bot^2/4],
$$
where ${\bf k}=({\bf k}_\bot,p_i)$, ${\bf k}_{0\bot}$ is time-
dependent, and  $\delta_l$ and $\delta_\bot$ - are longitudinal
and transversal packet widths (relative to the $z$ axis). This
representation follows from the study of electron states in a
magnetic field in the Appendix. For relativistic electrons,
the momentum uncertainty in the initial state is much less than 
the momentum proper. In real calculations of the numbers of 
emitted photons via (33), it is preferable to represent the
current operators as expansions in states with definite momentum 
at a given moment in time, with a time dependence characteristic 
of plane waves. In the present paper, this approximation is 
justified by the fact that due to the strong effect of the 
radiation on the particle's state in the comoving reference
frame, an effect exeeding the one produced by the external field,
we can ignore the quantization of levels in the time dependence
of the operators.Indeed, even the classical theory of synchrotron
radiation predicts that the mean energy of the photons emitted 
by a particle is much greater than $\omega_0$. In view of this, 
the mean difference in particle energies before and after photon
emission proves to be much greater than the separation between
the levels of transverse motion. Under these conditions, 
allowance for level quantization in the time dependence of the
operator can only lead to small corrections of order $1/{\overline n}$
(where  ${\overline n}\sim\gamma^3$ - is the mean ratio of the 
frequency of the emitted photon to  $\omega_0$). 

   As a result oof the action of electron operators, the vectors
$\bf k$ and $\bf k_1$ in the current operators in (33) are 
transformed into the vectors ${\bf k}_i-\Delta{\bf q}$ , where
$\Delta{\bf q}=\sum\limits_s{\bf q}_s$ , with  ${\bf q}_1,{\bf q}_2,...$ 
the momenta of emitted photons. Replacing the given expression
with ${\bf k}_i (t)={\bf k}_i-\Delta{\bf k}(t)$, where $\Delta{\bf k}(t)$
is the mean momentum loss by the particle by the time $t$, and
plugging it into all the cofactors in (33) that are not in the 
exponential, we get
\begin{equation}\label{34}
n_{\alpha\bf q}(t)={Z^2\over c^2}g_q^2\int\limits_0^t
dt_1\int\limits_0^t dt_2 e^{-i\omega(t_1-t_2)}\sum\limits_
{{\bf k}_i,\sigma^\prime}\vert C_{{\bf k}_i}\vert^2 ({\bf e}_
{\alpha\bf q}{\bf v}^*_{i\sigma^\prime}({\bf q},t_1))
({\bf e}^*_{\alpha\bf q}{\bf v}_{i\sigma^\prime}
({\bf q},t_2))*
$$
$$ 
\left(vac;{\bf k}_i,\sigma\vert{\widehat U}_0^\dagger(t_1)
{\widehat\rho}_{\bf q}^\dagger (t_1) {\widehat U}_0(t_1)
{\widehat U}_0^\dagger (t_2){\widehat\rho}_{\bf q} (t_2)
{\widehat U}_0(t_2)\vert {\bf k}_i,\sigma;vac\right),
\end{equation}
where
$$
{\widehat \rho}_{\bf q}(t)=\sum\limits_{{\bf k},\sigma,
\sigma^\prime} {\widehat d}^\dagger_{{\bf k}-{\bf q},\sigma^\prime}
{\widehat d}_{{\bf k},\sigma} {\rm exp}\lbrace i(\varepsilon_{\bf k-q}-
\varepsilon_{\bf k})t\rbrace ,
$$
\vskip 7pt
$$ 
v_{i\sigma^\prime}^a({\bf q},t)={c^2\over 2\sqrt
{\varepsilon_i\varepsilon_i^\prime}}w^*_{\sigma^\prime}\Biggl[
\sqrt{\varepsilon_i^\prime+mc^2\over {\varepsilon_i+mc^2}}
\sigma^a\sigma^b k_i^b(t)+\sqrt{\varepsilon_i+mc^2\over
\varepsilon_i^\prime+mc^2}\sigma^b(k_i^b(t)-
 q^b))\sigma^a\Biggr]w_{\sigma},
$$
with $\varepsilon_i=\varepsilon_{{\bf k}_i(t)},\hskip 5pt
\varepsilon_i^\prime=\varepsilon_{{\bf k}_i(t)-\bf q}$,
and the summation over repeated indices is implied. The
term coresponding to  $\sigma^\prime\neq\sigma$ describes
emission processes accompanied by electron spin flip.
   Further simplification is possible if the exponents in
the density operators in (34) are transformed according to
\begin{equation}\label{35}
\varepsilon_{\bf k-q}-\varepsilon_{\bf k}\approx
\sum\limits_s\mu ({\bf q},{\bf q}_s^\prime),
\end{equation}
where  $\mu ({\bf q},{\bf q}_s^\prime)$  are unspecified 
functions. In this approach, different photons are assumed 
to be  almost independent, since otherwise we would have to 
speak of a strong correlations between the emission of two 
separate photons, which agrees neither with the semiclassical
theory nor with the calculations below. In an approximation
that is linear in  $\Delta{\bf q}$, for  $q\ll k_i$ we have
\begin{equation}\label{36}  
\mu_i ({\bf q},{\bf q}_s^\prime) \approx (\nabla\varepsilon_
{\bf k_i-q}-\nabla\varepsilon_{\bf k_i}) {\bf q}_s^\prime 
\approx -{\bf q q}_s^\prime /m\gamma_i,
\end{equation}
where  $\gamma_i=\varepsilon_{{\bf k}_i}/mc^2$. 
As  $ q_s^\prime \to\infty $, the function  
$\mu ({\bf q}{\bf q}_s^\prime)$  ceases to depend on  $ q_s^\prime$.

   Using the methods of calculating means employed in 
Ref.9, we get
\begin{equation}\label{37}
n_{\alpha\bf q}(t)=\int\limits_0^t dt_1\int\limits
_0^t dt_2 \sum\limits_{{\bf k}_i,\sigma^\prime}\vert C_{{\bf k}_i}
\vert^2 {\dot Q}^*_{i\alpha\bf q}(t_1,\sigma^\prime){\dot Q}_
{i\alpha\bf q}(t_2,\sigma^\prime) {\rm exp}[-P_{i\bf q}(t_1,t_2)],
\end{equation}
where
$$ 
Q_{i\alpha\bf q}(t,\sigma^\prime)=i{Z\over c}g_q\int\limits_0^t
{\bf e}^*_{\alpha\bf q}{\bf v}_{i,\sigma^\prime}({\bf q},t^\prime) 
{\rm exp}[i\omega t^\prime-i{\bf qr}_{i,\sigma^\prime}({\bf q},
t^\prime)] dt^\prime,
$$
with $ {\dot{\bf r}}_{i,\sigma^\prime}({\bf q},t)={\bf v}_
{i,\sigma^\prime}({\bf q},t)$. The exponent in (37) is given by
\begin{equation}\label{38}
P_{i\bf q}(t_1,t_2)=\sum\limits_{\beta,{\bf q}^\prime,\sigma
^\prime}\biggl[\vert Q_{i\beta\bf q^\prime}(t_1,\sigma^\prime)
\vert^2\left(1-{\rm exp}[-i \mu_i({\bf q, q}^\prime) t_1]\right)+
$$
$$
\vert Q_{i\beta{\bf q}^\prime}(t_2,\sigma^\prime)\vert^2\left(1-
{\rm exp}[i\mu_i({\bf q, q}^\prime) t_2]\right)-
$$ 
$$ 
Q^*_{i\beta{\bf q}^\prime}(t_1,\sigma^\prime) Q_
{i\beta{\bf q}^\prime}(t_2,\sigma^\prime) \left(1-{\rm exp}[-i
\mu_i({\bf q, q}^\prime) t_1]\right)*
$$ 
$$
\left(1-{\rm exp}[i\mu_i({\bf q, q}^\prime) t_2]\right)\biggr].
\end{equation}
Obviously,   
$$
P^*_{i \bf q}(t_1,t_2)=P_{i \bf q}(t_2,t_1);
$$
$$
  \lim\limits_{q\to 0}P_{i \bf q}(t_1,t_2)\to 0;$$
$$  
\lim\limits_{t_1\to t_2}P_{i \bf q}(t_1,t_2)\to 0.
$$
Equation (37) contains the desired corrections to the semiclassical 
expression for the number of emitted photons. It assumes its 
semiclassical form for $\vert P_{i\bf q}(t_1,t_2)\vert \ll 1$.
From a physical standpoint, this difference betwwen the formulas
is due to the fact that in (37) we allow for interaction of the
emitted photons, while in the semiclassical theory this factor
is ignored. The probability distribution  for the number of
emitted photons in each state does not obey the Poisson law
any longer, which a reflection of the nonlinearity of
electromagnetic phenomena in the quantum theory.

   Obviously, an equation like (37) can be used to study
arbitrary motion of a particle, not just an electron in a 
synchrotron. To do so, we merely redefine the quantities
$ {\bf v}_{i\sigma^\prime}({\bf q},t)$, which in the simplex 
case can be approximately calculated for the mean of the
vector  ${\bf k}_i$ and averajed over spin (in this case, the
velocities  $ {\bf v}({\bf q},t)$  and the function (38)  no
longer depend on the indices  $i$ and $\sigma^\prime$).

   Let us estimate  $P_{\bf q}$  for the case in which the
velocity  $ {\bf v}({\bf q},t)$  is constant and equal to
${\bf v}_0$:
$$ 
Q_{\alpha\bf q}(t)={Z\over c} g_q {{\bf e}^*_{\alpha
\bf q}{\bf v}_0\over \omega-{\bf q v}_0} e^{i(\omega-{\bf q v}_0)t}.
$$
Plugging this into (38), we obtain an expression that is
logaritmically divergent, due to the slow decrease in the
integrands as  $ q^\prime\to\infty$. This fact is the
manifestation of ultraviolet divergence, often encountered 
in electrodynamics. In contrast to Feinman's perturbation 
theory, ultraviolet divergence does not lead to a catastrophe:
it only means that (in contrast to the predictions of the
semiclassical theory) a uniformly moving particleis not 
accompanied by transverse photons. This example is a clear 
demonstration of the dependence of the way in which the 
ultraviolet divergence depends on the perturbation theory
employed. A detailed study of this problem lies outside the 
scope of the present paper, where we use the standard method
of introducing a cutoff momentum  $ q_c\sim m c$ to remove
the ultraviolet singularity. The resulting expression for the
absolute value of the function (38) proves to be small and
varies very slowly (logaritmically) with $t_1$ and $t_2$. An
explicit estimate of the function (38) for $ {\bf v}({\bf q},t)$  
constant will be made in the next section.

\vskip 0.5cm
\noindent
{\bf 6. INFRARED ASYMPTOTIC BEHAVIOUR 
 OF THE NUMBER OF PHOTONS}

   Let us consider the asymptotic behaviour of the function
(37) as  $\omega=qc\to 0$. In classical electrodynamics (see, e.g.,
Ref. 11) and inÿ
 the semiclassical theory there is a characteristic
frequency dependence of  $ n_{\alpha\bf q}$ as $\omega\to 0$,
namely,  $n_{\alpha\bf q}\sim{1/\omega^3}$. Hence, upon integration
with respect to momenta, the total number of emitted photons
diverges logaritmically at the lower limit. Will allowing for the
effect of emission on a state of the emitting particle (as in Eq. 
(37)) influence this pattern? To ansver this question, we examing
a model problem in which a charged particle moves at constant 
velocity  ${\bf v}_1$ and, colliding at time $t_3>0$ with a point
scatterer, suddenly changes its own velocity by a small quantity
 $ \Delta{\bf v}={\bf v}_2-{\bf v}_1, \hskip 5pt \vert\Delta{\bf v}
\vert\ll v_1$, and then proceeds to move at constant velocity
 ${\bf v}_2$. The requirement that this jump in velocity be small
simplifies all calculations considerably. Moreover, since a jump 
in velocity implies infinite acceleration, various nonphysical
effects are to be expected. The requirement that the velocity 
jump be small makes the velocity almost a continuous function, so 
that such effects can be ignored. When Eq. (37) is employed in
calculations, there is the problem of the interaction suddenly 
turning on at the initial moment in time, which violates charge 
conservation, and of generation of fictitious radiation, which is 
the consequence of such violation. To avoid the need to discard 
fictitious terms, one can use the procedure developed in Sec. 3
to turn the interaction on slowly.
   
   Since the particle is assumed to have a definite velocity,
we drop the subscript $i$ in (38) and replace the vectors
${\bf v}_{i\sigma}$  with the current value of the velocity.
We calculate the resulting functions  $P_{\bf q}(t_1,t_2)$,
assuming that  $\tau_{in}\ll t_3\ll t$. To this end, we first 
estimate  the quantities  $ f(t_1,t_2)={\rm exp}\lbrack\pm i{\bf 
q q}^\prime t_{1,2}/m\gamma\rbrack$ on the right-hand  side of
Eq. (38). Since a photon is emitted when the particle changes
velocity, we consider the neighbourhood of the point  $t_1=t_3,
\hskip 4pt t_2=t_3$, assuming that  $t_3\sim\gamma_0 m/q^2$. 
The vector  $\bf q^\prime$  is the momentum transferred from
the moving particle to the emitted quanta of electromagnetic 
field (photons). The mean value of this momentum is of order
$m\vert\Delta{\bf v}\vert$, so that at $q\ll m\vert\Delta{\bf v}\vert$ 
the ratio  $q^\prime/ q$ can be large. Thus, the absolute
value of the exponent in  $f(t_1,t_2)$  in the range of 
parameters under investigation is large, and the exponentials
are rapidly varying functions that make a negligible contribution
to (38). Eliminating these contributions from the outset, we
reduce (38) to the simpler form
\begin{equation}\label{39}
P_{\bf q}(t_1,t_2)=\sum\limits_{\beta,{\bf q}^\prime}
\Bigl[\vert Q_{\beta\bf q^\prime}(t_1)\vert^2
+\vert Q_{\beta{\bf q}^\prime}(t_2)\vert^2-
$$
$$ 
Q^*_{\beta{\bf q}^\prime}(t_1) Q_{\beta{\bf q}^\prime}
(t_2) \left(1+e^{-i{\bf q q}^\prime (t_1-t_2)/m\gamma}\right) 
\Bigr].
\end{equation}
We now calculate the function (39) explicitly for
${\bf v}({\bf q},t)={\bf v}_0=const.$ In this case, assuming that 
$ q_c$  is much less than the mean momentum of the emitting particle,
we calculate the integral with respect to $q^\prime$ and obtain
\begin{equation} \label{40}
P_{\bf q}(t_1,t_2)={Z^2\over 4\pi^2 c^3}\int d o^\prime
{[{\bf n}^\prime\times {\bf v}_0]^2\over (1-{\bf n}^\prime
{\bf v}_0/c)^2} \biggl( i\, {\rm Si}(\omega_2 (t_1-t_2))+
$$
$$ 
i\, {\rm Si}((\omega_2+\omega_1)(t_1-t_2))+ 2 {\widetilde C}-
{\rm Ci}(\omega_2\vert t_1-t_2 \vert)-
$$
$$
{\rm Ci}(\vert\omega_2+\omega_1\vert\vert t_1-t_2 \vert)+
{\rm ln}\left( \omega_2 \vert\omega_2+\omega_1\vert 
(t_1-t_2)^2 \right)\biggr),
\end{equation} 
where $ {\bf n}^\prime={\bf q}^\prime/ q^\prime,\;
\omega_1=q_c {\bf n}^\prime {\bf q}/m\gamma,\; 
\omega_2=(c-{\bf n}^\prime {\bf v}_0) q_c, \;
{\rm Si}(\xi)$  and $\; {\rm Ci}(\xi)$ - are the sine and cosine
integrals, and  ${\widetilde C}=0.5772\cdots$ - is Euler's 
constant. The function (40) vanishes at $t_1 = t_2$ and slowly
increases with the time difference  $\Delta t=\vert t_1-t_2 \vert$. 
In the nonrelativistic limit at large  $\Delta t\gg 1/c q_c$ , the
function (40) can be arrpoximated by the expression
\begin{equation}\label{41}
P_{\bf q}(t_1,t_2)\approx {2 Z^2 v_0^2 \over 3\pi c^3}
\left[ i\pi\hskip 3pt {\rm sign}(t_1-t_2)+2\left({\widetilde C}+
{\rm ln}(c q_c)+{\rm ln}\vert t_1-t_2\vert \right)\right].
\end{equation}
We remark on the smallness of the coefficient of the expression
in square brackets. As $\Delta t$ increases, the real part of (41)
increases logarithmically, but the characteristic buildup time
proves to be exponentially large, so that the function (41) can be
considered small over the entire range of its arguments. 

   Now let us estimate the number of photons emitted by the electron
in the entire course of its motion for the nonrelativistic case.
Integrating by parts, we find, for instance, that   
$$ 
n_{\alpha\bf q}(t)=-i{Z^2\over c^2} g_q^2\int\limits_0^
t dt_1{\bf e}_{\alpha\bf q}{\bf v}(t_1) e^{-i\omega t_1+
i{\bf qr}(t_1)}*
$$
$$
\biggl[{{\bf e}^*_{\alpha\bf q}{\bf v}(t_2)\over{\omega-{\bf qv}
(t_2)+i\partial P_{\bf q}(t_1,t_2)/\partial t_2}}
\exp \lbrace i\omega t_2-i{\bf qr}(t_2)-P_{\bf q}(t_1,t_2)\rbrace
{\vrule width 0.4pt height 15pt depth 5pt \:}^{t_2=t}_{t_2=0}-
$$
$$
\int\limits_0^t dt_2 {\rm exp}\lbrace i\omega t_2-i{\bf qr}(t_2)-
P_{\bf q}(t_1,t_2)\rbrace
{\partial\over\partial t_2}\left({\bf e}^*_{\alpha\bf q}
{\bf v}(t_2)\over{\omega-{\bf qv}(t_2)+i\partial P_{\bf q}
(t_1,t_2)/\partial t_2}\right)\biggr].
$$
Allowance for the value of the first term inside the square brackets
at the lower limit is unjustified because of the violation of charge
conservation at $t\to 0$. If we turn the interaction on slowly, then
this contribution is zero. First we integrate by parts with respect
to $t_1$, using the same ideas that we used in integrating with 
respect to $t_2$. We obtain
\begin{equation}\label{42}
n_{\alpha\bf q}(t)\approx {Z^2\over c^2}g_q^2 \:
{\vrule width 0.6pt height 15pt depth 8pt \:}
{{\bf e}_{\alpha \bf q} {\bf v}_2 \over \omega -{\bf q v}_2
+i\partial P_{\bf q}(t,t_2)/ \partial t_2}
{\vrule width 0.4pt height 10pt depth 15pt \:}_{t_2=t}\:
{\vrule width 0.6pt height 15pt depth 8pt \:}^2 +
$$
$$
{Z^2\over c^2}g_q^2 \int
\limits_0^t dt_2 \int\limits_0^t dt_1 \exp\lbrace i\omega (t_2-t_1)+
i{\bf q}({\bf r}(t_1)-{\bf r}(t_2))-P_{\bf q}(t_1,t_2)\rbrace*
$$
$$
{\partial\over\partial t_1}\biggl[{{\bf e}_{\alpha\bf q}{\bf v}(t_1)
\over\omega-{\bf qv}(t_1)-i\partial P_{\bf q}/\partial t_1}\;
{\partial\over\partial t_2}\left({{\bf e}^*_{\alpha\bf q}
{\bf v}(t_2)\over\omega-{\bf qv}(t_2)+i\partial P_{\bf q}/
\partial t_2}\right)\biggr],
\end{equation}
\vskip 4pt\noindent
where we have discarded the rapidly oscillating terms, which contribute
nothing to the overall expression for the number of emitted photons.
The first term on the right-hand side of Eq.(42) corresponds to the 
part of the transverse field that follows the moving particle, and is
related neither to change in the particle's velocity nor to the
radiation. Hence in all calculations of the characteristics of the 
radiation that follow, we allow only for the second (integral) term.

In calculating the time derivatives in (42) we encounter continuous 
and delta-function terms, with the latter being a reflection of the
discontinuity in velocity, the derivatives  ${\dot Q}_{\alpha
\bf q}$ and ${\dot P}_{\bf q}$. For instance,
\begin{equation}\label{43}
{\partial\over\partial t_2}\left({{\bf e}^*_{\alpha\bf q}
{\bf v}(t_2)\over\omega-{\bf q}{\bf v}(t_2)+
i\partial P_{\bf q}(t_1,t_2)/\partial t_2}\right)=
$$
$$
 i\Theta(t_3-t_2){\partial^2 P_{\bf q}(t_1,t_2)\over
\partial t_2^2}\; {{\bf e}^*_{\alpha\bf q}
{\bf v}_1\over(\omega-{\bf q}{\bf v}_1+
i\partial P_{\bf q}(t_1,t_2)/\partial t_2)^2}+
$$
$$ 
i\Theta(t_2-t_3){\partial^2 P_{\bf q}(t_1,t_2)\over
\partial t_2^2}\hskip 4pt {{\bf e}^*_{\alpha\bf q}
{\bf v}_2\over(\omega-{\bf q}{\bf v}_2+
i\partial P_{\bf q}(t_1,t_2)/\partial t_2)^2}+
$$
$$ 
\delta(t_2-t_3)\biggl[
{{\bf e}^*_{\alpha\bf q}{\bf v}_2\over\omega-{\bf q}
{\bf v}_2+i\partial P_{\bf q}(t_1,t_2)/\partial t_2}
{\vrule width 0.4pt height 10pt depth 15pt\,}
_{t_2=t_3+0}-
$$
$$
{{\bf e}^*_{\alpha\bf q}{\bf v}_1\over\omega-{\bf q}
{\bf v}_1+i\partial P_{\bf q}(t_1,t_2)/\partial t_2}
{\vrule width 0.4pt height 10pt depth 15pt\,}
_{t_2=t_3-0}\biggr].
\end{equation}
Here $\Theta(\xi)$  is the Heaviside step function.

The relationship between the continuous and delta-function
terms in (43) can be evaluated as follows. The total contribution
of the $\Theta$-functions can again be calculated by parts, which
again results in a delta-function contribution multiplied by the 
magnitude of the discontinuity of the integrand at $t_2=t_3$.
This jump includes the second derivative of $P_{\bf q}$ as a factor
whose order of magnitude can be estimated to be the product of the
first derivative and the mean value of the frequency of the emitted 
photon. The latter cannot exceed the energy lost by the moving particle,
and it is therefore proportional to the small parameter 
$\lambda={\bf v}_1\Delta{\bf v}/v_1^2$. Clearly, allowing for the 
continuous terms in (43) would mean allowing for the next terms in the
series expansion of the integrals in $\lambda$. the leading term is
still the contribution of the delta function, the only contribution
we consider.

   Using the condition that the interactionis turned on slowly, we  find
that
$$ 
Q_{\alpha\bf q}(t)={Z\over c}g_q \; 
{{\bf e}^*_{\alpha\bf q}{\bf v}_1\over\omega-{\bf qv}_1}
e^{i(\omega -{\bf qv}_1)t},\; \tau_{in}\ll t\le t_3.
$$
For $t_3<t$ the result is different:
$$ 
Q_{\alpha\bf q}(t)={Z\over c}g_q\biggl[
\left({{\bf e}^*_{\alpha\bf q}{\bf v}_1\over\omega-{\bf qv}_1}-
{{\bf e}^*_{\alpha\bf q}{\bf v}_2\over\omega-{\bf qv}_2}\right)
e^{i(\omega-{\bf qv}_1)t_3}+
{{\bf e}^*_{\alpha\bf q}{\bf v}_2\over\omega-{\bf qv}_2}
e^{i(\omega -{\bf qv}_2)t}\biggr].
$$
For  $\tau_{in}\ll t_2\le t_3$ we have
$$
{\partial P_{\bf q}(t_1,t_2)\over\partial t_2}=-i\sum\limits
_{\beta{\bf q}^\prime} Q^*_{\beta{\bf q}^\prime}(t_1)\biggl[{Z\over
c}g_{q^\prime}({\bf e^*_{\beta\bf q^\prime} v}_1) e^{i(\omega-
{\bf qv}_1)t_2}\bigl(1+
$$
$$ 
\exp \lbrace-i{\bf qq}^\prime (t_1-t_2)/m\gamma\rbrace\bigr)+
{{\bf qq}^\prime\over m\gamma} Q_{\beta{\bf q}^\prime}(t_2)
\exp \lbrace-i{\bf qq}^\prime (t_1-t_2)/m\gamma\rbrace
\biggr].
$$
Finally, for  $t_2>t_3$ we have
$$
{\partial P_{\bf q}(t_1,t_2)\over\partial t_2}=\sum\limits
_{\beta {\bf q}^\prime}\biggl[ -2{Z^2\over c^2} g^2_{q^\prime}
\sin\lbrace(\omega^\prime-{\bf q}^\prime{\bf v}_2){t_2-t_3\over 2}
\rbrace ({\bf e^*_{\beta{\bf q}^\prime}v}_2)\biggl({{\bf e}_
{\beta{\bf q}^\prime} {\bf v}_1\over
{\omega^\prime-{\bf q}^\prime{\bf v}_1}}-
$$
$$
{{\bf e}_{\beta{\bf q}^\prime} {\bf v}_2\over{\omega
^\prime-{\bf q}^\prime{\bf v}_2}}\biggr)-
i{{\bf qq}^\prime\over m\gamma} Q^*_{\beta{\bf q}^\prime}(t_1)
Q_{\beta{\bf q}^\prime}(t_2) \exp \lbrace-i{\bf qq}^\prime(t_1-
t_2)/m\gamma\rbrace-
$$
$$ 
iQ^*_{\beta{\bf q}\prime}
(t_1) {Z\over c} g_{q^\prime}({\bf e}^*_{\beta{\bf q}^\prime}
{\bf v}_2)e^{i(\omega^\prime-{\bf q}^\prime{\bf v}_2)t_2}\bigl(1+
\exp \lbrace-i{\bf qq}^\prime(t_1-t_2)/m\gamma\rbrace\bigr)
\biggr].
$$
Note that  çâ® $\partial P_{\bf q}(t_1,t_2)/\partial t_2$ is
continuous in $t_1$. 

   Let us calculate the delta-function  contribution  to  the  integrals
with respect  to $t_2$ in  (42), letting  $t\to\infty$ :
$$ 
n_{\alpha\bf q}(\infty)={Z^2\over c^2}g_q^2 \int
\limits_0^\infty dt_1 \exp \bigl[-i\omega(t_1-t_3)+i{\bf q}
({\bf r}_0(t_1)-{\bf r}_0(t_3))-P_{\bf q}(t_1,t_3)\bigr]*
$$
$$
{\partial\over\partial t_1}\biggl[{{\bf e}_{\alpha\bf q}
{\bf v}(t_1)\over\omega-{\bf qv}(t_1)-i\partial P_{\bf q}
(t_1,t_3)/\partial t_1}\biggl({{\bf e}^*_{\alpha\bf q}{\bf v}_2
\over\omega-{\bf q}{\bf v}_2+i\partial P_{\bf q}(t_1,t_2)/
\partial t_2}{\vrule width 0.4pt height 10pt depth 15pt\,}
_{t_2=t_3+0}-
$$
$$
{{\bf e}^*_{\alpha\bf q}{\bf v}_1\over\omega-{\bf q}
{\bf v}_1+i\partial P_{\bf q}(t_1,t_2)/\partial t_2}
{\vrule width 0.4pt height 10pt depth 15pt\hskip 2pt}
_{t_2=t_3-0}\biggr)\biggr].
$$
Now we integrate with respect to  $t_1$,  again  limiting  ourselves  
to delta-function contributions. Allowing for the fact that 
 $P_{\bf q}(t_2,t_1)=P_{\bf q}^*(t_1,t_2)$, we obtain
\begin{equation}\label{44}
n_{\alpha\bf q}(\infty)={Z^2\over c^2}g_q^2 
\; \vrule width 0.4pt height 15pt depth 15pt
\; {{\bf e}^*_{\alpha\bf q}{\bf v}_2
\over\omega-{\bf q}{\bf v}_2+i\partial P_{\bf q}(t_3,t_2)/
\partial t_2}{\vrule width 0.4pt height 10pt depth 15pt\,}
_{t_2=t_3+0}-
$$
$$ 
{{\bf e}^*_{\alpha\bf q}{\bf v}_1\over\omega-{\bf q}
{\bf v}_1+i\partial P_{\bf q}(t_3,t_2)/\partial t_2}
{\vrule width 0.4pt height 10pt depth 15pt\hskip 2pt}
_{t_2=t_3-0}\hskip 4pt{\vrule width 0.4pt height 15pt depth 
15pt \hskip 4pt}^2.
\end{equation}
This equation solves the problem. If we neglect the derivatives of
$P_{\bf q}$ in  the  denominators,  (44)  coincides  with  the  standard
expression for  the  number  of  low-frequency  photons  emitted  in   a
collision,  the  expression   that   can   be   derived   in   classical
electrodynamics$^{11}$ and in quantum electrodynamics if we  use  standard
perturbation theory$^6$. Let us estimate the derivatives of $P_{\bf  q}$
in the denominators in (44). We have
$$ 
{\partial P_{\bf q}(t_3,t_2)\over\partial t_2}
{\;\vrule width 0.4pt height 10pt depth 15pt\;}
_{t_2=t_3+0}=-2i{Z^2\over c^2}\sum\limits_{{\bf q}^\prime}
g^2_{q^\prime} {[{{\bf q}^\prime}\times{\bf v}_1]
[{{\bf q}^\prime}\times{\bf v}_2]\over {q^\prime}^2(\omega^
\prime-{{\bf q}^\prime}{\bf v}_1)}+O_1(q),
$$
where  $O_1(q)$ is of the first order in $q$ .  For  $q$  small,  noting
that 
$$ 
{\partial P_{\bf q}(t_3,t_2)\over\partial t_2}
{\; \vrule width 0.4pt height 10pt depth 15pt\; }
_{t_2=t_3-0}=-2i{Z^2\over c^2}\sum\limits_{{\bf q}^\prime}
g^2_{q^\prime} {[{{\bf q}^\prime}\times{\bf v}_1]^2
\over {q^\prime}^2(\omega^\prime-{{\bf q}^\prime}{\bf v}_1)}+
O_1(q), 
$$
and that  $ {\bf v}_2\approx {\bf v}_1$, we obtain
\begin{equation}\label{45} 
n_{\alpha\bf q}(\infty)={Z^2\over c^2} g_q^2 
\hskip 4pt\vrule width 0.4pt height 15pt depth 15pt\;
{{\bf e}_{\alpha\bf q}{\bf v}_2\over\omega-{\bf q}{\bf v}_2+\Delta}-
{{\bf e}_{\alpha\bf q}{\bf v}_1\over\omega-{\bf q}
{\bf v}_1+\Delta}\; {\vrule width 0.4pt height 15pt depth 
15pt} \; ,
\end{equation}
where
\begin{equation}\label{46}
\Delta=2 {Z^2\over c^2}\sum\limits_{{\bf q}^\prime}
g^2_{q^\prime}{[{{\bf q}^\prime}\times{\bf v}_1]^2
\over {q^\prime}^2(\omega^\prime-{{\bf q}^\prime}{\bf v}_1)}.
\end{equation}
In the nonrelativistic limit  $v_0\ll c$ , from (46) we obtain
$\Delta\approx 4 Z^2 v^2_1 q_c/3\pi c^2$. Equation (45) does not contain
the infrared singularity. A reviation from the $ n_{\alpha\bf q}\sim 
1/\omega^3$ law with decreasing $\omega$ begins at an  energy  of  order
$\Delta$. The lower  the  energy  of  relative  motion  of  the  charged
particle and point scatterer, the lowe the aforementioned  energy.  This
estimate also holds if the velocity of the particle changed not suddenly
but over a  time  interval  that  is  short  compared  to  the  time  of
production of a low energy photon. 

\vskip 0.5cm

\noindent
{\bf 7. CALCULATING THE DENSITY MATRIX FOR THE CASE}
{\bf OF SYNCHROTRON RADIATION}
 
   We now use the above approach to calculate the density martrix of  an
emitting particle. The exact expression for the density  matrix  in  the
representation realized by the transformation (23) has the form   
\begin{equation}\label{47}
\gamma ({\bf x},{\bf x}^\prime,t)=<t\vert{\widetilde
{\psi}}^\dagger ({\bf x},t){\widetilde{\psi}}({\bf x}^\prime,t)
\vert t>.
\end{equation}
We calculate (47) in the first approximation, replacing the vector $|t>$
by the initial state vector  $\vert 0>$. Using the Baker-Hausdorff  rule
with  proper  transformation  of  the  evolution  operators   ${\widehat
U}_0(t)$ and ${\widehat U}_0^\dagger (t)$, and the operators 
${\widehat Q}_{\alpha\bf q}(t)$  in the form (20), we easily find that
\begin{equation}\label{48}
\gamma ({\bf x},{\bf x}^\prime,t)=\gamma_0({\bf x},{\bf x}^
\prime,t)\, \exp [-S({\bf x-x^\prime},t)],
\end{equation}
where  $\bf \gamma_0({\bf x},{\bf x}^\prime,t)=\phi^*({\bf x},t)
\phi({\bf x}^\prime,t)$ is the value of the density  matrix  tthat  does
not account for emission and is determined by the wave function 
$\phi({\bf x},t)$ of the exactly described state of the electron  in  an
external magnetic field.  The function  $S({\bf x-x^\prime},t)$  in  the
exponent is given by
\begin{equation}\label{49}
S({\bf x-x^\prime},t)=\sum\limits_{{\bf q},\alpha}{\vert
Q_{\alpha\bf q}(t)\vert}^2 [1-e^{i{\bf q(x-x^\prime})}] 
\end{equation}
which vanishes at $ \bf {x=x^\prime}$. As $ \vert{\bf x-x^\prime}\vert\to
\infty $, the function (49) acquires the maximum  value,  equal  to  the
total number of photons emitted by the given moment in time. 

   The mean momentum of the particle can be evaluated as follows:
\begin{equation}\label{50}
{\overline{\bf p}(t)}={\bf p_0}+i\int \vert\phi({\bf x},
t)\vert^2\nabla^\prime S({\bf x},{\bf x}^\prime,t)\vert_
{{\bf x=x}^\prime}\; d^3 x.
\end{equation}
This means that the gradient $\nabla^\prime S({\bf x},{\bf x}^\prime,t)$ 
determines the rate of decrease of the mean  particle  momentum  due  to
emission of photons. If the initial state was stationary, $\vert\phi({\bf 
x},t)\vert^2$ does not depend on time. In  this  case,  the  mean  force
acting on the particle is
\begin{equation}\label{51}
{\bf F}_b=i\int\vert\psi_0({\bf x},t)\vert^2\nabla^\prime
\dot{S}({\bf x},{\bf x}^\prime,t)\vert_{{\bf x}^\prime={\bf x}}\;
d^3 x.
\end{equation}
The calculation of the function $ S({\bf x},{\bf x}^\prime,t)$  for  the
case of synchrotron radiation is  similar  to  the  calculation  of  the
proton production rate in Sec.4. Noting that $S$ actually depends on the
difference  ${\bf r}={\bf x}-{\bf x}^\prime$, we obtain the value of $s$
averaged over one period:
\begin{equation}\label{52}
\overline{S}({\bf r},t)=t{Z^2\over c}\int\limits_0^\pi d\theta 
\sin\theta \sum\limits_{n=1}^\infty n\omega_0\biggl[{\rm cot}^2\theta\:
J_n^2({nv_0\over c}\sin\theta)+
$$
$$
{{v_0^2}\over{c^2}}{J_n^\prime}^2({nv_0 \over c}\sin\theta)\biggr]
\left(1-J_0(r{{n\omega_0}\over c}\sin\theta_0\:\sin\theta)\: \
\exp (ir{{n\omega_0}\over c}\cos\theta_0\: \cos\theta)\right),
\end{equation}
\vskip 5pt\noindent
where  $\theta_0$  is the polar angle of  the  vector   ${\bf  r}$  with
respect to the axis perpendicular to the orbital plane, $ r=\vert{\bf r}
\vert$. In the ultrarelativistic case the following approximate formula
is more convenient:
\begin{equation}\label{53}
{\overline S}({\bf r},t)=t{2^{2/3}Z^2\omega_0\over c}\int
\limits_{\varsigma_0}^\infty d\varsigma\varsigma^{1/3}
\int\limits_{\theta_-}^{\theta_+} d\theta \sin\theta *
$$
$$
\Bigl\lbrace {\rm ctg}^2\theta \:{\rm Ai}^2[(\varsigma/2)^{2/3}
(1-{v_0^2\over c^2}\sin^2\theta)]+
$$
$$
{v_0^4\over c^4}{2^{2/3}\sin^2\theta\over\varsigma^{2/3}}\:{{\rm Ai}^\prime}^2
[(\varsigma/2)^{2/3}(1-{v_0^2\over c^2}\sin^2\theta)]\Bigr\rbrace *
$$
$$
\left(1-J_0(\sin\theta_0 \:\sin\theta\: {r\varsigma\omega_0\over
c})\:
\exp(i \cos\theta_0 \:\cos\theta\: {r\varsigma\omega_0\over c})\right),
\end{equation}
\vskip 5pt\noindent
where $\varsigma_0=\epsilon^{-3},\; \theta_- = \pi/2 - \epsilon,\;
\theta_+ = \pi/2 + \epsilon$ and $1/\gamma\ll\epsilon\ll 1$;
${\rm Ai} (z)$ is the Airy function defined in the Ref.12 and 
${\rm Ai}^\prime (z)$ is its derivative. Obviously, the imaginary part 
of the averaged expression for $S$ given by (52) and (53) is zero.

   According to Ref.10, the density matrix (48) describes an ensemble of
pure states  (in  the  sense  of  von  Newmann),  whose  properties  are
determined by the behaviour of  $ e^{-S({\bf r},t)}$. The  expansion  of
the matrix (48) in the density matrices of  these  pure  states  can  be
written
\begin{equation}\label{54}
\gamma ({\bf x,x}^\prime ,t)=\int d^3a\Phi^*_{\bf a}({\bf x},
t)\hskip 2pt \Phi_{\bf a}({\bf x}^\prime,t)\hskip 2pt N_
{\bf a}(t),
\end{equation}
where  $ \Phi_{\bf a}$ are the wave functions of the  pure  states,  and
$ N_{\bf  a}(t)d^3a$   are  the  probabilities  that  these  states  are
realized at the given momentum in time. The functions $\Phi_{\bf a}({\bf 
x},t)$ are proportional to the products of the wave function $\phi  {\bf
x},t)$  and the wave functions  $ \chi({\bf x-a},t)$, where 
$ \chi ({\bf x},t)$  is the solution of the integral equation
\begin{equation}\label{55}
G({\bf r},t)=e^{-S({\bf x-x}^\prime,t)}=\int\chi^*({\bf x-a},t)
\, \chi ({\bf x}^\prime-{\bf a},t)\hskip 2pt d^3a.
\end{equation}
But what about the existence and uniqueness  of  the  solution  of  this
equation? If we write (55) in the Fourier representation
$$
G_{-\bf q}(t)=\chi^*_{\bf q}(t)\chi_{\bf q}(t),
$$
the absolute value of the desired function is uniquely defined, but  not
the phase. However, this uncertainty is  a  direct  consequence  of  the
translation invariance of Eq.(55), whose  general  solution,  therefore,
has the form
\begin{equation}\label{56}
\chi_{\bf q}=e^{i\alpha}\sqrt{G_{-\bf q}}
\end{equation}
with  arbitrary  real   $\alpha$.  A  solution  exists  if  the  Fourier
transform  $G_{\bf q}$  is a real nonnegative quantity. That it is real
follows directly from the fact that  ${\rm Im} S({\bf r},t)$ is odd and 
${\rm Re} S({\bf r},t)$ even under  insversion;  the  nonnegativity  
follows from the fact that  ${\rm Re} S({\bf  r},t)$  increases  
monotonically  with distance $r$.

   The effective size of the localization region for the  initial  state
in  the  orbital  plane  is  $\delta\rho\sim\sqrt{R/\gamma  v_0}$   (see
Appendix). The quantity  $\delta\rho$ is usually much larger than atomic
dimensions. The localization region for  the  initial  state  along  the
magnetic field  is  infinitely  large,  which  is  due  to  the  initial
uncertainty  in  the  $z$-component  of  the  momentum.  The  latter  is
obviously  determined  by  the  macroscopic  parameters  of  the  actual
experimental layout. 

   Radiation can substantially alter the picture, and  lead  to  spatial
localization of the emitting particle in a region whose size is  of  the
order of atomic dimensions. To estimate the rate  of  variation  of  the
widths of the states  $ \Phi_{\bf a}(t)$ with the passage of  time,  the
function  $ S({\bf r},t)$  was  calculated  numerically  for  a  set  of
parameters characteristic of the FIAN--60 synchrotron ($E = 0.68 GeV$ and
$R = 2 m$).
\vskip 5pt
\noindent
{\bf 8. MAIN CONCLUSIONS}

   The perturbation theory developed in this paper has made it  possible
to   establish   that   certain   fundamental   problems   of    quantum
electrogynamics are not invariant when the type of  perturbation  theory
is altered. With respect to infrared divergence, this is shown by emloying
the simple example of an emitting particle that undergoes a sudden change 
in velocity. The results have been obtained for the nonrelativistic case,
since the study of emitting relativistic particles requires a detailed 
analysis  of the ultraviolet asymptotic behavior of the integrand in (38).

   The method of removing ultrfviolet divergences by introducing a cutoff 
momentum, which was adopted in the present paper, is not covariant under
Lorenz transformation, and therefore cannot be used in a consistent 
relativistic theory. But even preliminary studies show that in the new
approach the problem of ultraviolet divergence is not catastrophic, in
contrast to the case in ordinary perturbation theory of quantum 
electrodynamics. It is to be hoped that further research will lead to
progress in understanding this problem.

   Density matrix calculation have shown that reduction of the spatial 
dimensions of the localization region for the emitting electrons to atomic 
dimensions can be achieved over a macroscopically long time interval
$\tau_c$ of some tenths of a second. Can the present theory be applied to 
such long times? The situation is complicated by the fact that  in the 
course of one orbital revolution, the particle is subject to a solenoidal
electric field that balanced the loss of energy to photon emission. If we 
assume that this field acts during a time interval so short that it only
accelerates the particle's wave packet as a whole and is unable to change 
the particle's internal parameters substantially, then there is no reason 
why to do estimates we cannot extend the theory to the entire duration of
the particle's motion in the synchrotron. 

   The time $\tau_c$ is much shorter than it takes the packet to spread 
due to the nonequidistant nature of the spectrum of the transverse-motion 
levels. What is observed is an anisotropy in the packet's width: the packet 
is most strongly squeezed perpendicular to the magnetic field, and least 
strongly parallel to the field. The considerable elongation of the packet 
in the direction of the magnetic field is obvious.

   The posibility of strong spatial localization of the emitting particles
means that if the acceleration cycle in the synchrotron is long enough,
the motion of the particle can be described to high accuracy by the
equations of classical mechanics. Nevertheless, this does not mean that 
the intensity of the radiation must agree with the prediction of classical
electrodynamics. Indeed, a localized state in quantum mechanics is
completely different in its properties from a localized state in Newton's
classical theory. The justification for using Newton's equations of motion
to calculate the paths followed by wave packets is provided by Ehrenfest's
theorem, but the decisive factor in calculating the intensity of the 
radiation is the momentum of the particle, rather than the position. In 
quantum mechanics, a state with a definite momentum is completely 
delocalized, and in this way differs substantially from states of type
 $ \Phi_{\bf a}$ . There is thus no way in which we can intuitively
interpret calculations of the characteristics of radiation using
classical ideas. The characteristic common feature of the formulas
derived in this paper is the fact that allowing for the mutual 
interaction of the emitted photons reduces the radiative intensity.
A similar result was obtained by Landau and Pomeranchuk$^{13}$, who
studied the radiation emitted by charged particles moving in continuous 
media (the Landau---Pomeranchuk effect). The physics of this phenomenon 
amount to the fact that random collisions of an emitting particle with
particles of the medium can reduce the path length over which the
radiative indensity builds up coherently. Something similar is observed
when photons are emitted into vacuum: multiple emission of photons can
mimic the multiple collisions in a continuous medium that lead to a
reduction in radiative intensity.
  
   The numerical calculations were done together with V.A.Aleksandrov.
The work was supported by a grant from the Russian Fund for Basic 
Research (Grant No. 97-02-16058).

\vskip 0.5cm
\noindent
{\bf APPENDIX: COHERENT STATES OF A RELATIVISTIC}

\noindent
{\bf ELECTRON IN A UNIFORM MAGNETIC FIELD}

  Let 
$$
{\bf A}=(-{1\over 2}y{\rm H}_0,{1\over 2}x{\rm H}_0,0),
$$ 
where  ${\rm H}_0$  is the strength of the magnetic field directed along
the $z$ axis. The motion of an electron in such a field, which obeys the
Dirac equation, has been the topic of numerous studies (see, e.g., Refs.
14-16). The solution given below differs from the well-known one only in
some details. 

   We introduce the lowering operators for the two independent oscillators:
$$
{\widehat a}_1={1\over 2}\sqrt{m\omega_L}(x+i{\widehat p}_x/m\omega_L-
iy+{\widehat p}_y/m\omega_L);
$$
$$
{\widehat a}_2={1\over 2}\sqrt{m\omega_L}(x+i{\widehat p}_x/m\omega_L+
iy-{\widehat p}_y/m\omega_L),
$$
where  $\omega_L=\vert e\vert {\rm H}_0/2mc$ is the Larmor frequency. The
frequency of the first independent oscillator is twice the Lormor frequency,
$\omega_1=2\omega_L$, while the frequency of the second oscillator is zero.
The set of lowering and raising operators (which are conjugates of 
lowering operators) satisfies the standard Bose commutation relations.
The operators  ${\widehat a}_1$ and ${\widehat a_1}^\dagger$ describe the
orbital motion of an electron in a magnetic field, while the operators
${\widehat a}_2$ and ${\widehat a}_2^\dagger$ describe the position, 
fluctuations and other characteristics of the center of the osculating
circular orbit, whose mean radius is $R$. 

   We next introduce the matrix operator
$$
{\widehat D}=
\left(\matrix{
{\widehat p}_z & -2i\sqrt{m\omega_L}\hskip 3pt {\widehat a}_1 \cr
2i\sqrt{m\omega_L}\hskip 3pt {\widehat a}_1^\dagger & -{\widehat p}_z \cr
}\right).
$$
The energies of the electron's quantum states are
$$
E_\xi=\sqrt{m^2c^4+p^2 c^2+4\omega_L mc^2(n_1+\sigma+{1\over 2})}
,\eqno(A.1)
$$
where the label $\xi=(n_1,n_2,\sigma,p)$ simply indicates the set of quantum
numbers in parentheses. The  $n_1,n_2 \vert=0,1,2,...$ label the quantum 
states of the independent oscillators, with  $n_1$  being the principal
quantum number. We denote the projection of momentum on the $z$ axis by
$p$. The discrete variable $\sigma$ takes two values,  $\pm 1/2$ , 
corresponding to two possible projections of spin on the direction of the
magnetic field. The bispinor describing a stationary state of an electron
in a magnetic field is given by
$$
\psi_{\xi}({\bf r},t)={1\over\sqrt{2 E_\xi}}
\left(\matrix{
\sqrt{E_{\xi}+mc^2}\varphi_{\xi}({\bf r}) \cr
{c\over\sqrt{E_{\xi}+mc^2}}{\widehat D}\varphi_{\xi}({\bf r}) \cr
}\right) e^{-iE_{\xi}t},\eqno(A.2)
$$ 
where  $\varphi_{\xi}({\bf r})$  is a spinor of the form
$$
\varphi_{\xi}({\bf r})=e^{ipz}{1\over \sqrt{n_1 ! n_2 !}}
({\widehat a}_1^\dagger)^{n_1}({\widehat a}_2^\dagger)^{n_2}
\varphi_{0\sigma}({\rho}).\eqno(A.3)
$$
Here 
$$
\varphi_{0\sigma}({\rho})=\sqrt{m\omega_L/\pi}\:
\exp (-m\omega_L\rho^2/2) \hskip 2pt \chi_{\sigma},
$$
with 
$$
\rho^2=x^2+y^2, \;\chi_{1/2}^*=(1,0),\; \chi_{-1/2}^*=(0,1),
$$
and the normalization length along the $z$ axis is taken equal to unity.

   An arbitrary solution of the Dirac equation is a linear combination of
bispinors of type (A2). Just what linear combination corresponds to the
initial state of an electron in the synchrotron? For standard values of
synchrotron parameters ( for example, for the FIAN--60 synchrotron), the 
mean value of  $n_1$ is very large (of order $10^{13}$), and if the
assumption that photons are emitted largely independently of one another
is true, so is Poisson's law. In this case, the expected  value  of  the
relative fluctuation of the number $n_1$ is extremely small, $\lambda=
{\overline{\Delta n_1}}/{\overline n_1}\sim 10^{-6}$. Hence, essentially
all terms of the desired linear combination can be expanded in powers of
 $\lambda$, with the result that
$$
E_{\xi}\approx E_{\sigma p}+(2\omega_L/\gamma_{\sigma p})
\Delta n_1,
$$
where  $E_{\sigma p}$  is the value of  $E_\xi$  at $\xi=({\overline n_1},
{\overline n_2},\sigma,p)$, and $\gamma_{\sigma  p}=E_{\sigma  p}/mc^2$.
We see that the spectrum  is  essentially  uniformly  spaced,  with  the
leveles being separated by the mean orbital frequency $\omega_{\sigma p}=
2\omega_L/\gamma_{\sigma p}$ of the electron about the  magnetic  fiald.
When the relative fluctuation of $n_1$ is small, we can put $E_{\xi}
\approx E_{\sigma p}$ in all nonexponential factors. 

   The linear combination corresponding to the  above  requirements  has
the form
$$
\psi_{\sigma p}({\bf r},t)={1\over\sqrt{2 E_{\sigma p}}}
\left(\matrix{
\sqrt{E_{\sigma p}+mc^2}\varphi_{\sigma p}({\bf r},t) \cr
{c\over\sqrt{E_{\sigma p}+mc^2}}{\widehat D}\varphi_{\sigma p}
({\bf r},t) \cr
}\right) \exp (-i\Delta E_{\sigma p}t),\eqno(A.4)
$$ 
where  $\Delta E_{\sigma p}=E_{\sigma p}-\omega_{\sigma p}{\overline n_1}$,
and
$$
\varphi_{\sigma p}({\bf r},t)=e^{ipz} \exp \bigl[\sqrt
{\overline n_1}( e^{i\alpha_1}{\widehat a}^\dagger_1(t)-e^
{-i\alpha_1}{\widehat a}_1(t)) +
$$
$$ 
\sqrt{\overline n_2} ( e^{i\alpha_2}{\widehat a}^\dagger_2(t) - 
e^{-i\alpha_2}{\widehat a}_2(t))\bigr] \varphi_{0\sigma}({\rho}),
$$
with  ${\widehat a}_1^\dagger (t)={\widehat a}_1^\dagger \exp (i\omega_
{\sigma p} t)$, and ${\widehat a}_2^\dagger (t)={\widehat a}_2^\dagger $, 
where ${\alpha}_1$ and ${\alpha}_2$ are constant  phases;  the  momentum
along  the $z$ axis is assumed equal to $p$. The components of the current 
density vector in the state (A4) are
$$
j_{\sigma p}^{x}={2c^2\over E_{\sigma p}} \sqrt{m\omega_L 
{\overline n}_1} \vert\varphi_{\sigma p}({\bf r},t)\vert^2
\sin(\omega_{\sigma p}t+\alpha_1)
$$
$$
j_{\sigma p}^{y}=-{2c^2\over E_{\sigma p}} \sqrt{m\omega_L 
{\overline n}_1} \vert\varphi_{\sigma p}({\bf r},t)\vert^2
\cos(\omega_{\sigma p}t+\alpha_1)\eqno(A.5)
$$
$$
j_{\sigma p}^{(z)}=-{c^2 p\over E_{\sigma p}}\vert\varphi_{\sigma p}
({\bf r},t)\vert^2.
$$

   The packet's rms width in the radial direction in the state (A4) is
determined by the radial behavior of the function 
$\varphi_{0\sigma}({\rho})$ and can be estimated to be $\Delta\rho=
\sqrt{2Rc/E_{\sigma p}}$. In the azimutal direction, the stationary states
of type (A2) are completely delocalized. Indeed, in these states the
angular momentum is well-defined, and by virtue of the uncertainty
relation for action-angle variables, they cannot be localized in angle.

   In contrast, the state (A4) has no definite angular momentum, but its
angular width is limited, and is of order  $\Delta\phi\sim 1/\sqrt{
{\overline n}_1}$ in the azimutal direction (we assume that the uncertainty
in the position of the orbit's center is much smaller than the orbit's 
radius, so that  $n_2\ll n_1$), which after being multiplied by the orbit's
radius yields a distance roughly equal to  $\Delta\rho$ (for the FIAN-60
synchrotron this distance is about one micrometer). 

   The packet width along the $z$ axis is governed by such macroscopic 
parameters of the device as the diaphragm width, and for this reason it
can exceed the radial or azimutal width many times over. In this case the 
packet can be represented by a linear combination of states of type (A4):
$$
\psi_{\sigma}({\bf r},t)=\sum\limits_p C_p \psi_{\sigma p}({\bf r},
t),\eqno(A.6)
$$
where the constants  $C_p$  satisfy the normalization condition and 
guarantee, e.g., a Gausian dependence on the $z$ projection of the momentum
with midpoint at  $p=0$:
$$
C_p=(2\pi \delta_0^2)^{1/4} e^{-p^2\delta_0^2/4}.
$$
If we assume that the spatial width of the packet along the $z$ axes is
much greater thanthe radial width, then $\delta_0 \gg \Delta\rho $, and in
this case the state (A6) is associated with a small symmetric ellipsoid
elongated in the direction of the magnetic field and revolving in this
orientation ina circular orbit about an axis parallel to $z$. To estimate 
the time of packet spread in the radial or azimutal direction, we must keep 
the next term in the expansion of the energy  $E_\xi$ in powers of 
 $\Delta n_1$. This yields the value of the time of packet spreading due
to the nonequidistant levels of transverse motion,  $\tau_1 \sim \gamma_0 
R^2/{\overline{\Delta n_1}}$. Here $\gamma_0$ is the Lorenz factor for the
electron beam in a synchrotron. For the FIAN--60 synchrotron the time
$\tau_1$  was estimated to be about ten seconds.

\end{document}